\documentclass{elsart}

\usepackage{natbib}

 \usepackage{graphics}
 \usepackage{graphicx}
 \usepackage{epsfig}

\usepackage{amssymb}

\begin{document}

\begin{frontmatter}



\title{ Dynamical model of financial markets: 
fluctuating `temperature' causes intermittent behavior of 
price changes}

 \author[label1]{Naoki Kozuki}, 
 \author[label2]{Nobuko Fuchikami\corauthref{cor1}}
 \ead{fuchi@phys.metro-u.ac.jp} 
 \corauth[cor1]{Correponding author.}
 \address[label1]{Hitachi High-Technologies Corporation, 
 24-14, Nishi-shimbashi, 1-chome, Minato-ku, Tokyo 105-8717, Japan}
 \address[label2]{Department of Physics, 
 Tokyo Metropolitan University, Minami-Ohsawa, 
 Hachioji, Tokyo 192-0397, Japan}



\begin{abstract}
We present a model of financial markets 
originally proposed for a turbulent flow, 
as a dynamic basis of its intermittent behavior.  
Time evolution of the price change is 
assumed to be described by Brownian motion in a power-law potential,
where the `temperature' fluctuates slowly. 
The model generally yields a fat-tailed distribution of the price change. 
Specifically a Tsallis distribution 
is obtained if the inverse 
temperature is  $\chi^{2}$-distributed, which qualitatively agrees with 
intraday data of foreign exchange market. 
The so-called `volatility', 
a quantity indicating the risk or activity in financial markets, 
corresponds to the temperature of markets and its fluctuation leads to 
intermittency.

\end{abstract}


\end{frontmatter}

\section{Introduction}
\label{introduction}

Financial returns are known to be non-gaussian and exhibit 
fat-tailed distribution \cite{fa65}\nocite
{ms00,bg01,gbp96,ms96,abc96,ms97,ams98,fpr00,bgt00,go97,mdd97,ms95,
lgc99}-\cite{hkk01}. 
The fat tail relates to intermittency --- 
an unexpected high probability of large price changes, which is of 
utmost importance for risk analysis. 
The recent development of high-frequency data bases makes it possible to 
study the intermittent 
market dynamics on time scales of less than a day 
\cite{ms00}\nocite{
bg01,gbp96,ms96,abc96,ms97,ams98,fpr00,bgt00,go97,mdd97,ms95,
lgc99}-\cite{hkk01}. 

Using foreign exchange (FX) intraday data, M\"uller et al. \cite{mdd97} 
showed that 
there is a net flow of information from long to short timescales, i.e., 
the behavior of long-term traders influences the behavior of short-term 
traders. 
Motivated by this hierarchical structure, Ghashghaie et al. \cite{gbp96} have 
discussed analogies between the market dynamics and hydrodynamic 
turbulence \cite{fr95,cgh90}, and claimed that the information cascade 
in time hierarchy exists in a FX market, which corresponds to the energy 
cascade in space hierarchy in a three-dimensional turbulent flow. 
These studies have stimulated further investigations on similarities 
and differences in statistical properties of the fluctuations in the 
economic data and turbulence 
\cite{ms96}\nocite{abc96,ms97,ams98,fpr00}-\cite{bgt00}. 
Differences have also emerged. 
Mantegra and Stanley \cite{ms96,ms97} and Arneodo et al. \cite{abc96} 
pointed out that the time evolution, or equivalently the power spectrum 
is different for the price difference (nearly white spectrum) and 
the velocity difference ($f^{1/3}$ spectrum, i.e., $-5/3$-law 
for the spectrum of the velocity). 
Moreover, from a parallel analysis of the price change data with 
time delay and the velocity difference data with time delay 
(equivalent to the velocity difference data with spatial 
separation under the Taylor hypothesis \cite{fr95}), it was shown that 
the time evolution of the second moment and the shape of the probability 
density function (PDF), i.e., the deviation from gaussian PDF are 
different in these two stochastic processes \cite{ms97}. 
On the other hand, non-gaussian character in fully developed 
turbulence \cite{fr95} has been linked with the nonextensive statistical 
physics \cite{ts88}\nocite{ww00,aa01,be01,be01a}-\cite{bls01}. 

As dynamical foundation of nonextensive statistics, Beck recently 
proposed a new model describing hydrodynamic turbulence \cite{be01,be01a}. 
The velocity difference $\Delta v$ of two points in a turbulent flow 
with the spatial separation $\Delta r$ is described by Brownian motion 
(an overdamped Langevin equation \cite{ri89}) in a power-law potential. 
Assuming a $\chi^{2}$-distribution for the inverse temperature, 
he obtained a Tsallis distribution \cite{ts88} for $\Delta v$. 
However, if we take into account the almost uncorrelated behavior of 
the price change \cite{ms96}\nocite{abc96}-\cite{ms97}, 
the picture by means of the Brownian motion seems to be more appropriate 
for the market data rather than turbulence. 
Moreover, the description by the Langevin equation is able to relate 
the PDF of the price change to that of the volatility, 
a quantity known as a measure of the risk in the market. 
Thus we applied the model to FX market dynamics by employing the 
correspondence by Ghashghaie et al. \cite{gbp96}.

\section{Model}
\label{model}

We substitute the FX price difference 
$Z(t) \equiv y(t+\Delta t)-y(t)$ with the time delay $\Delta t$ for 
the velocity difference $\Delta v$ with the spatial separation $\Delta r$. 
Beck's model for turbulence then reads
 
\begin{equation}
\frac{dZ}{dt} = \gamma F(Z)+R(t),
\label{dZdt}
\end{equation}
where $\gamma > 0$ is a constant, and $R(t)$ is gaussian white noise 
corresponding to the temperature $kT$, satisfying 
$\left< R(t)R(t') \right> = 2\gamma kT\delta (t-t')$. 
The `force' $F=-\partial U(Z)/\partial Z$ is assumed to be 
obtained by a power-law potential $U(Z)=C|Z|^{2\alpha}$ with an 
exponent $2\alpha$, where $0<\alpha \leq 1$ and $C$ is a positive constant. 
That is, the system is subject to a restoring force proportional to the power 
of the price difference, $|Z|^{2\alpha}$  besides the random force $R(t)$. 
Especially when $\alpha =1$, the restoring force is linear to $Z$. 
Under a constant temperature $kT$, Eq. (\ref{dZdt}) leads to a 
stationary (i.e., thermal equilibrium) distribution of $Z$ as 

\begin{equation}
P_{\Delta t}(Z|\beta)  = 
 \frac{e^{-\beta U(Z)}}{\int e^{-\beta U(Z)}dZ} = 
 \frac{\alpha}{\Gamma(\frac{1}{2\alpha})}(C\beta)^{1/2\alpha}
 e^{-\beta C|Z|^{2\alpha}},
\label{PZbeta} 
\end{equation}

where $\beta \equiv 1/kT$ is the inverse temperature \cite{be01}. 
The `local' variance of $Z$, which is defined for a fixed value of $\beta$, 
is obtained from the conditional probability in Eq. (\ref{PZbeta}) as
\begin{equation}
\left< Z^{2} \right>_{\beta} = 
 \int Z^{2}P_{\Delta t}(Z|\beta)dZ =  
 \frac{\Gamma(\frac{3}{2\alpha})}{\Gamma(\frac{1}{2\alpha})}
 (C\beta)^{-1/\alpha} \propto{(kT)}^{1/\alpha}.
 \label{Z2beta}
\end{equation}

We define the volatility by the square root of the local variance of $Z$ 
(see Eq. (\ref{V})). 
When $\alpha = 1$ and $C=1/2$, $\left<Z^{2}\right>_\beta$ coincides with 
the temperature and the conditional PDF reduces to gaussian, 
while for $\alpha \neq 1$, 
$\{
\left<
Z^2
\right>_{\beta}
\}^{\alpha}$ 
is proportional to $kT$. 

Let us assume that the `temperature' of the FX market is not constant and 
fluctuates in larger time scales, and $\beta$ is, just as in Beck's model 
for turbulence, $\chi^{2}$-distributed with degree $n$:

\begin{equation}
f_{\Delta t}(\beta ) \equiv \frac{1}{\Gamma(\frac{n}{2})}
\left( 
\frac{n}{2\beta _0} 
\right)^{n/2}
\beta^{n/2-1} 
\exp\left(
-\frac{n\beta}{2\beta_0}
\right),
\qquad n>2\,, 
\label{fbeta}
\end{equation}
where $\Gamma$ is the Gamma function, $\beta_{0}$ is the average of 
the fluctuating $\beta$ and $n$ relates to the relative variance of $\beta$:

\begin{equation}
\left< \beta\right>=\beta_0, \qquad
\frac{\left< \beta^2 \right> - \left< \beta \right>^2}
     {\left< \beta \right>^2} = \frac{2}{n}\,.
\label{beta}     
\end{equation}
Equation (\ref{fbeta}) implies that the local variance 
$\left< Z^{2}\right>_{\beta}\equiv v$ fluctuates with the distribution 
$p(v)\propto v^{-(\alpha+1)}f_{\Delta t}(v^{-\alpha})$. 
The conditional probability in Eq. (\ref{PZbeta}) together with 
Eq. (\ref{fbeta}) yields a Tsallis-type distribution \cite{ts88,be01}  
for the ultimate PDF of $Z$:

\begin{eqnarray}
P_{\Delta t}(Z)  &=& 
\int P_{\Delta t}(Z|\beta)f_{\Delta t}(\beta) d\beta = 
\frac{1}{Z_q} 
\frac{1}
     { 
     \left\{
     1+\frac{ C \beta_0 2 \alpha (q-1) |Z|^{2\alpha} }  {2 \alpha -(q-1)}      
       \right\}^{1/(q-1)}
       }\;, 
\label{PZ}       \\
\frac{1}{Z_q} &=& \alpha \left\{
\frac{C \beta_0 2 \alpha (q-1)}{2 \alpha-(q-1)}
\right\}^{1/2\alpha} 
\frac{\Gamma\left( \frac{1}{q-1} \right)}
     {\Gamma(\frac{1}{2\alpha}) 
     \Gamma(\frac{1}{q-1}-\frac{1}{2\alpha})}\,,
\label{Zq}
\end{eqnarray}
where Tsallis' nonextensivity parameter $q$ is defined by 

\begin{equation}
q \equiv 1+ \frac{2 \alpha}{\alpha n+1}\,,
\label{q}
\end{equation}
which satisfies $1<q < 5/3$ because of $0<\alpha \leq 1$ 
and $n>2$. 
Since $q>1$, the distribution of $Z$ exhibits power-law tails for 
large $Z$: 
$P_{\Delta t}(Z)\sim Z^{-2\alpha/(q-1)}=Z^{-(\alpha n+1)}$. 
Hence, the $m$th moment $\left< Z^{m}\right >=\int Z^{m}P_{\Delta t}(Z)dZ$ 
converges only for $m<\alpha n$. 
In the limit of $q\rightarrow 1$, $P_{\Delta t}(Z)$ in Eq. (\ref{PZ}) 
reduces to the canonical distribution of extensive statistical mechanics: 
$\lim_{q\rightarrow 1}P_{\Delta t}(Z)=P_{\Delta t}(Z|\beta_{0})$.

\section{Results and discussions}
\label{results}

We have applied the present model to the same FX market data set as 
used in Ref. \cite{gbp96} (provided by Olsen and Associates \cite{oa93} which consists of 
1 472 241 bid-ask quotes for US dollar-German mark exchange rates during 
the period October 92 - September 93). 
The volatility is often estimated by the standard deviation of the price 
change in an appropriate time window \cite{ms00}. 
Employing this definition, we express the volatility in terms of 
the local standard deviation of the price change 
$Z(t_i) \equiv y(t_i+\Delta t)-y(t_i)$ as 

\begin{equation}
V\equiv \sqrt{
\frac1N \sum_{j=1}^N \left\{Z(t_{j})\right\}^{2}
}\,.
\label{V}
\end{equation}
Here the window size has been chosen as $N=35$. 
Since $V^{-2\alpha}$ corresponds to $\beta$ (see Eq. (\ref{Z2beta})) 
which is $\chi^{2}$-distributed, $n$ can be explicitly obtained from 
the relative variance of $\beta$ using Eq. (\ref{beta}). 
Thus, there is only one adjustable parameter among $(\alpha,q,n)$, 
because we have another relation, Eq. (\ref{q}). 
In other words, the PDF, $P_{\Delta t}(Z)$ of the price change and 
the PDF, $f_{\Delta t}(\beta)$ of the inverse power 
$V^{-2\alpha} \equiv \beta$ of the volatility are determined simultaneously 
once the value of $\alpha$ has been specified. 

\begin{figure}[htbp] 
\centering{\resizebox{9cm}{!}{\includegraphics{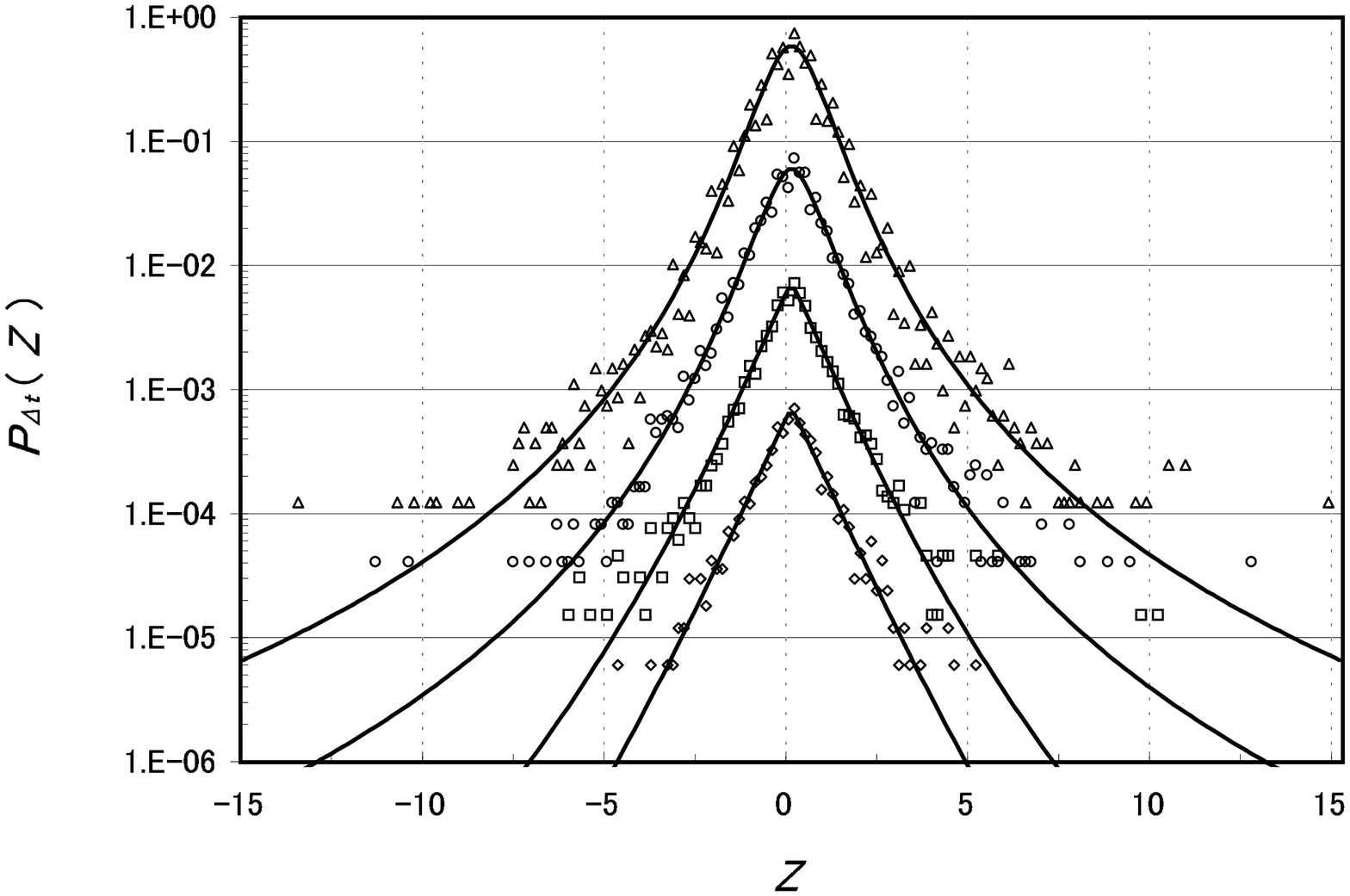}}}
\caption{
Data points: standardized PDF $P_{\Delta t}(Z)$ of price changes 
$Z(t) \equiv y(t+\Delta t)-y(t)$ for time delays $\Delta t =$ 320 s,  
1 280 s,  5 120 s,  20 480 s (from top to bottom). 
The middle prices $y(t) \equiv (y_{\rm bid}(t) + y_{\rm ask}(t))/2$ have 
been used (data provided by Olsen and Associates \cite{oa93}). 
The PDF has been obtained in a similar way as Ref. \cite{ms95}, i.e., 
we have selected the complete set of non-overlapping records 
separated by a time interval $\Delta t (1 \pm \epsilon)$ with the 
tolerance $0< \epsilon <0.035$. 
The number of the available data points thus decreases with $\Delta t$, 
which is 54 087, 16 338, 4 359, 1 114 (from top to bottom). 
For better visibly, the curves have been vertically shifted with 
respect to each other.
Full lines: theoretical expression given in Eq. (\ref{PZ}) in text, 
where $(\alpha, q, n)$ is (0.918, 1.393, 4), (0.832, 1.322, 5), 
(0.565, 1.101, 18), (0.540, 1.054, 35) (from top to bottom).  
$n$ is explicitly decided from the variance of the volatility data 
(the closest integer satisfying Eq. (\ref{beta})), $q$ is from 
Eq. (\ref{q}), so that we only need to adjust $\alpha$ by least 
square fitting.  
}
\label{Tsallis}
\end{figure}

\begin{figure}[htbp] 
\centering{\resizebox{12cm}{!}{\includegraphics{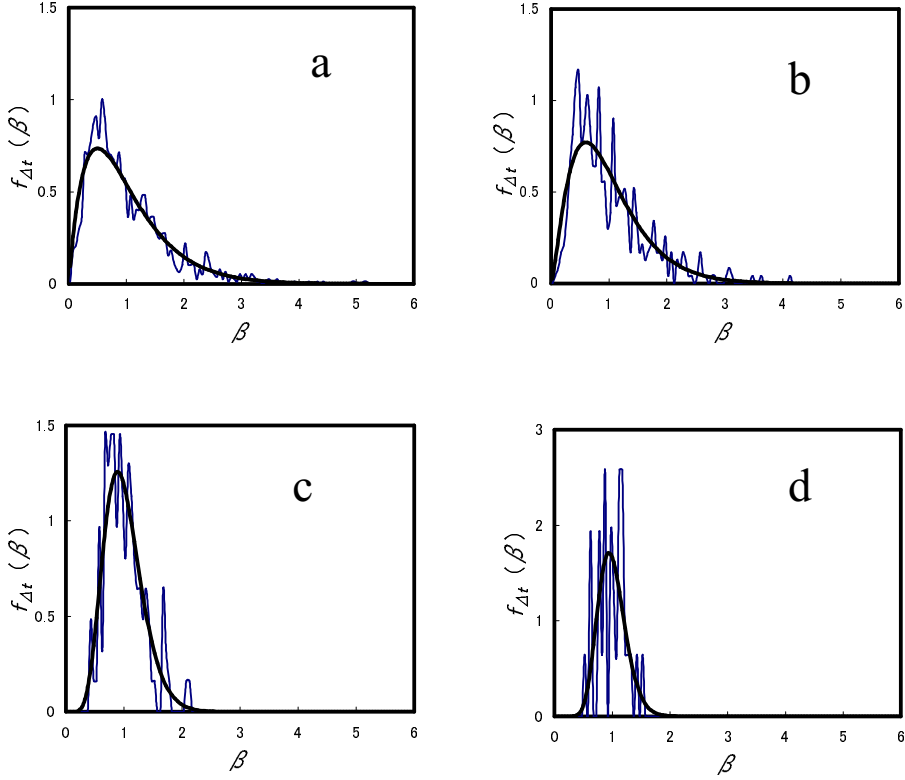}}}
\caption{
Rescaled probability density $f_{\Delta t}(\beta)$ of 
the quantity $V^{-2\alpha}\equiv \beta$, 
where $V$ is the volatility defined by the local standard deviation of 
the price change $Z(t)$. 
(See the legend of Fig. \ref{Tsallis} for the definition of $Z(t)$.) 
Thin lines: obtained from the volatility data, which was defined by 
$V(t_i) \equiv \sqrt{
\sum_{j=1}^N \left\{
                   Z(t_{N(i-1)+j}) 
                   \right\}^2/N
}$ 
with $N=35$ so that the windows for different volatility data points 
do not overlap. 
The number of the available data points is 1 545, 466, 124, 24 (from a to d). 
The data set has been rescaled so that the average of $\beta$ is unity: 
$\left< \beta \right>= \left< V^{-2\alpha}\right> = 1$. 
Thick lines: $\chi^2$-distribution for $\beta$ given in 
Eq. (\ref{fbeta}) with $\beta_0 =1$. 
The same parameter values as in Fig. 1 are used, 
where $n$ is the integer closest to $2/\left( \left< \beta^2\right>-1\right)$ 
(see Eq. (\ref{beta}) in text), $\alpha$ has been adjusted by 
least square fitting of $P_{\Delta t}(Z)$ in Fig. 1. 
($\alpha, q, n$)  is (0.918, 1.393, 4), (0.832, 1.322, 5), 
(0.565, 1.101, 18), (0.540, 1.054, 35) (from a to d).
}
\label{x2}
\end{figure}

The PDF's with time delay $\Delta t$ varying from five minutes up to 
approximately six hours are displayed in Fig.\ref{Tsallis} 
together with theoretical 
curves obtained from Eq. (\ref{PZ}). 
As the time scale $\Delta t$ increases, $n$ increases, 
while $\alpha$ and $q$ decrease. 
The nonextensivity parameter $q$ tends to the extensive limit: 
$q \rightarrow 1$ as $\Delta t$ increases. 
Using the same parameter values $(\alpha,q,n)$, the PDF's of $\beta$ are 
compared in Fig. \ref{x2}. 
The average `temperature' in the market 
$1/\beta_{0} \sim \left< Z^{2} \right>^{\alpha} 
\sim (\Delta t)^{\alpha \zeta_{2}}$ 
increases with $\Delta t$ since $\alpha \zeta_{2}$ is positive. 
(We obtained the scaling exponent \cite{gbp96,fr95} $\zeta \approx 0.85$, 
which is larger than 2/3 obtained for turbulence.) 
In contrast, the fluctuation of the temperature increases with decreasing 
$\Delta t$ because the variance of the inverse temperature is proportional 
to $2/n$. 
The smaller values of $n$ imply the stronger intermittency 
which occurs in small time scales. 
The intermittent character in the price 
change can be seen as a fat tail of $P_{\Delta t}(Z)$ in Fig. \ref{Tsallis}. 
Also in Fig. \ref{x2}, the peak of $f_{\Delta t}(\beta)$ shifts to 
smaller $\beta$ as $\Delta t$ decreases (from d to a in Fig. \ref{x2}), 
which means relatively high temperatures are realized more frequently 
in short time scales.

It should be noted that the PDF in Eq. (\ref{PZ}) with $\alpha =1$ reduces 
to Student's  $t$-distribution \cite{ms00}, 
which has often been used to characterize the fat tails \cite{hkk01}. 
When $\alpha =1$, there is no adjustable parameter because $n$ is decided 
from Eq. (\ref{beta}). 
The FX market system is then subject to a restoring force linear to 
the price change and the volatility is proportional to the temperature. 
We have found that the PDFs of $Z$ and $\beta$ reproduce, 
although very roughly, the Olsen and Associates' data points 
even if $\alpha$ is fixed at $\alpha =1$. 
However, adjusting the parameter $\alpha$ improved the line shape of 
$P_{\Delta t}(Z)$ in a range close to $Z=0$. 
The data points in Fig. \ref{Tsallis} exhibit a cusp at $Z=0$ for large 
time scales $\Delta t$, which implies a singularity of the second 
derivative of the PDF at $Z=0$. 
The larger reduction of $\alpha$ from unity leads to the stronger 
singularity. 
(Note that the factor $|Z|^{2\alpha-2}$ arises from 
$d^2 P_{\Delta t}(Z)/ d Z^2$.) 
Thus the better fitting for large $\Delta t$ was obtained from a 
reduced value of $\alpha$. 
The trend of decrease in $\alpha$ with increasing $\Delta t$ was 
observed for the turbulent flow \cite{be01} as well. 
However, the deviation from $\alpha=1$ is much smaller than the present 
case and the cusp is invisible. 

Ghashghaie et al. \cite{gbp96} have used a model for turbulence by 
Castaing et al. \cite{cgh90}, in which a log-normal distribution has been 
assumed for the local standard deviation of the price change. 
The present model reduces to the model by Ghashghaie et al. 
if the stochastic process given in Eq. (\ref{dZdt}) is assumed 
with $\alpha=1$ (the local variance of $Z$ is then proportional to $kT$) 
and the $\chi^2$-distribution for $\beta$ 
(the inverse of the local variance of $Z$) is replaced by 
the log-normal distribution for $\sqrt{kT}$ 
(the local standard deviation of $Z$). 
Although no analytic expression like Eq. (\ref{PZ}) for $P_{\Delta t}(Z)$ 
is obtained, a similar qualitative explanation can be applied to their model: 
The volatility (or equivalently, the square root of the temperature) 
fluctuates slowly with a log-normal distribution, 
and the smaller time scale corresponds to the larger variance of the 
logarithm of the volatility. 
(The variance is denoted by $\lambda^2$ in Ref. 4.) 
However, the power-law behavior of the tail of the volatility 
distribution \cite{lgc99} can be better described by the $\chi^2$-distribution 
for the inverse of the variance. 

We have proposed the stochastic process described by Eq. (\ref{dZdt}) 
for FX market dynamics in small time scales. 
In fact, Eq. (\ref{dZdt}) is the simplest stochastic process 
which can realize the thermal equilibrium distribution, Eq. (\ref{PZbeta}) in 
the power-law potential. 
More realistic and more complicated processes that assure 
convergence to Eq. (\ref{PZbeta}) at local temperatures might be possible. 
Mantegna and Stanley have proposed a different stochastic model of 
the price change, which is described by a truncated Levy flight 
(TLF) \cite{ms95,ms00}. 
The model, yielding approximately a stable distribution, well reproduces 
the self-similar property of PDF at different time scales:  
$\Delta t =$ 1 min to 1 000 min. 
However, the parameters $\alpha$ and $\gamma$ characterizing the stable 
distribution fluctuate for larger (monthly) time scales \cite{ms00}, 
where $\gamma$ gives a measure of the volatility. 
In other words, the ultimate distribution 
(let us denote it $P^{\rm TLF}_{\Delta t}(Z)$) should be obtained, 
like Eq. (\ref{PZ}), from the weighted average over these parameters. 
A difference between $P^{\rm TLF}_{\Delta t}(Z)$ and $P_{\Delta t}(Z)$ is 
that $P^{\rm TLF}_{\Delta t}(Z)$ has no cusp at $Z=0$: 
$P^{\rm TLF}_{\Delta t}(Z) \sim 1-AZ^2$ for $|Z| \ll 1$, 
whereas $P_{\Delta t}(Z) \sim 1-A|Z|^{2\alpha}$, and the present FX date 
set indeed exhibits a cusp as seen in Fig. \ref{Tsallis}. 

Finally, a fundamental question has been left open: 
How to derive theoretically the increasing trend of fat-tailed character 
with decreasing time scale, i.e., $\Delta t$-dependence of 
the parameters ($\alpha, q, n$). 
An attempt to derive the non-gaussian fat-tailed character in small time 
scales was made by Friedlich et al. recently \cite{fpr00}. 
They derived a multiplicative Langevin equation from a Fokker-Planck 
equation and showed that the equation becomes more multiplicative 
and hence fat-tailed as $\Delta t$ decreases. 
Clarifying the relation between their multiplicative Langevin 
equation and Eq. (\ref{dZdt}) (the latter being rather simple although 
including the fluctuating temperature) should be important as well as 
interesting. 

{\bf Acknowledgments}


We would like to thank Professor T. Watanabe for valuable comments. 
The FX data set was provided by Olsen and Associates.




\end{document}